\documentclass[aps,prl,twocolumn,superscriptaddress,nofootinbib,floatfix,
nobalancelastpage,nobibnotes]{revtex4}
\usepackage{epsfig}
\usepackage{graphicx}
\usepackage{caption}

\begin{document}
\title
{Neutrino collective effects and sterile neutrino production in the early universe}

\author{R. F. Sawyer}
\affiliation{Department of Physics, University of California at
Santa Barbara, Santa Barbara, California 93106}

\begin{abstract}
Neutrino-neutrino interactions, as mediated by standard model Z exchange, can drastically enhance the
Dodelson-Widrow mechanism for the production of sterile neutrinos in the region of temperature $5-15$ MeV. For a sterile mass in the KeV region with the usual type of active-sterile neutrino coupling, the production can be copious enough to provide ample dark matter, even when the active-sterile coupling is small enough to avoid the bounds set by laboratory results.   
\end{abstract}
\maketitle
 
 \subsection{1. The underlying theory. }
 Recently it has been shown, quite in contrast with previous expectations, that in the early universe in the period
 T=1-10 MeV, the individual neutrinos in the seas are executing rapid flavor gyrations \cite{RFS1}. Their time scale is many orders 
of magnitude smaller than both the expansion time scales for the universe and that for a neutrino to scatter off of anything in the particle bath (or to be emitted or absorbed).
 The phenomenon has nothing to do with ordinary $\nu$ oscillations. It rests on the $\nu-\nu$ force
 transmitted by Z meson exchange in the standard model of weak interactions. 
 
Our neutrinos are combinations of plane wave states in a periodic box of side $t_{\rm max}$, the time necessary in order to see our phenomena, of the order of 1mm/c in what follows. In the 3D case the 
 $10^{30}$ or so mode momenta  $p_j$ for each flavor are drawn randomly from this set of states, in the present case from a Fermi distribution with zero chemical potential and definite temperature T. The probability that any particular momentum $p_1$ matches any other momentum from the draw is essentially zero.
 
 Much of the discussion below is limited to an example in what I will call 1(+) dimensions. Ref. \cite{RFS1}
 gives most of the required technical foundation, and explores the connections to 3D as well. As to 1(+)D, we start by defining a ``beam" as a set of neutrinos all of the same flavor, and very nearly in the same direction but with a tiny angular spread in a cone of angle $\Delta \theta$. The entire initial physical state consists of an assemblage of these units. But for now in the 1(+)D case application we envision having two sets of these beams, centered in opposing (R and L) directions. If we subdivided even further with many such cones occupying the same tiny solid angle and allowed them to overlap as much as they please, the calculation must yield the same results. 
 
 The answer to ``Why not have taken these states in precisely the two central directions rather than in these cones; they can be of any absolute momentum without the dynamics caring, as it turns out, and in the end we are going to cover the sphere with an angular mesh in any case?"  is that the statistical mechanics that sets the initial ensemble cares a lot.

We introduce flavor operators for a two flavor system, $\vec\sigma^j $'s for the N different $\nu_j$'s in the R beam, and   $\vec\tau^j $'s for those in the L beam; and use the notation $\sigma_+=\sigma_1+i \sigma_2$, etc. We write an effective Hamiltonian coming from Z exchange,
 \cite{rs}, but here omitting the usual bilinear neutrino oscillation term, and ``$\nu$-potentials coming from coherent interactions with other particles in the sea", which are both irrelevant to the physics at hand,

  \begin{eqnarray}
&H_{\rm eff}={G_F\over {\rm Vol.} }\sum_{j,k}\Bigr [\sigma_+^j\tau_-^k +\sigma_-^j\tau_+^k 
+{1\over 2}\sigma_3^j \tau_3^k +
\bar \sigma_+^j \bar \tau_-^k +\bar\sigma^j_- \bar\tau^k_+
\nonumber\\
&+{1 \over 2} \bar \sigma^j_3 \bar\tau^k_3 
 -\bar \sigma^j_+\tau^k_- - \bar \sigma^j_-\tau^k_+ -\sigma_+^j\bar \tau_-^k -\sigma_-^j \bar \tau_+^k
 \nonumber\\
&-( \bar \sigma_3^j \tau_3^k -  \sigma_3^j \bar \tau_3^k )/2
\Bigr ] (1-\cos \theta_{j,k})\,.
\nonumber\\
\label{hammy}
\end{eqnarray}

The (j,k) indices enumerate both the individual particle momenta and their initial flavors. In (\ref{hammy}) we take $\cos \theta=-1$  between all pairs of states in which one comes from a  $\vec \sigma$ beam and one from a $\vec\tau$ beam. Now the individual momenta do not enter in any way into our 1(+)D picture. Therefore we can sum over the momentum states $\vec p$ of each beam and redefine collective $\sigma,\tau$ operators, 
\begin{eqnarray}
\sum_j^N \vec \sigma^j \rightarrow \sqrt N \vec \sigma^j~,~\sum_j^N \vec \tau^j \rightarrow \sqrt N \vec \tau^j~,
\label{coll}
\end{eqnarray}
The redefined operators and their mates for $ \bar \sigma, \bar \tau $ obey the commutation rules of Pauli matrices. The factor of $N$ that emerged
combines with the ${\rm Vol.} ^{-1}$ factor in $H_{\rm eff}$ to reset the all-over coupling constant in (\ref{hammy})
be $G_F n$, where $n$ is the particle density. We choose time units so that this factor is unity. Now, with the momenta
out of the way, what remains of the indices $j,k$ refer only to the flavors of the initial states in the early universe mix. 
The equations of motion are,

\begin{eqnarray}
i {d\over dt} \sigma_+^j=\sum_k^4\Bigr [\sigma_3^j( \tau_+^k -\bar \tau_+^k)+ \sigma_+^j(\bar \tau_3^k-\tau_3^k)\Bigr]~,
\nonumber\\
i {d\over dt}\tau_+^j=\sum_k^4\Bigr[\tau_3^j( \sigma_+^k -\bar \sigma_+^k)+ \tau_+^j(\bar \sigma_3^k-\sigma_3^k)\Bigr]~,
\nonumber\\
i {d\over dt}\bar \sigma_+^j=\sum_k^4 \Bigr [\bar \sigma_3^j( \bar \tau_+^k - \tau_+^k)+\bar\sigma_+^j( \bar\tau_3^k-\tau_3^k)\Bigr]~,
\nonumber\\
i {d\over dt} \bar\tau_+^j=\sum_k^4 \Bigr [\bar \tau_3^j (\bar \sigma_+^k - \sigma_+^k)+\bar \tau_+^j( \bar \sigma_3^k- \sigma_3^k)\Bigr ] ~,
\label{array1}
\end{eqnarray}
and,
\begin{eqnarray}
i {d\over dt} \sigma_3^j=2\sum_k^4\Bigr [(\sigma_+^j \tau_-^k - \sigma_-^j \tau_+^k)-(\sigma_+^j \bar \tau_-^k
 - \sigma_- ^j \bar\tau_+^k)\Bigr ]~,
\nonumber\\
i {d\over dt} \tau_3^j=-2\sum_k^4\Bigr [(\sigma_+^k \tau_- ^j - \sigma_-^k \tau_+^j)-(\bar \sigma_-^k  \tau_+^j -   \bar\sigma_+^k \tau_-^j)\Bigr ]~,
\nonumber\\
i {d\over dt} \bar \sigma_3^j=2\sum_k^4\Bigr [(\bar\sigma_-^j\bar\tau_+^k- \bar \sigma_+^j \bar\tau_-^k)+
(\bar \sigma_-^j  \tau_+^k - \bar\sigma_+^j \tau_-^k)\Bigr]~,
\nonumber\\
i {d\over dt} \bar \tau_3^j=-2\sum_k^4\Bigr [(\bar \sigma_+^k\bar  \tau_-^j -  \bar\sigma_-^k \bar \tau_+^j)+
(\sigma_+^k \bar \tau_-^j - \sigma_-^k \bar\tau_+^j)\Bigr]~.
\label{array2}
\end{eqnarray}
 
where the range $k=1,..4$ is sufficient to describe the 1(+)D universe as we construct it. We take an initial state with separate beams in which the operators $[\sigma_3,\bar \sigma_3, \tau_3, \bar \tau_3]$ take the respective values $[\pm 1, \pm 1,\pm 1, \pm 1]$; 16 beams in all.
 The solutions that inspired the present work are those of ref.'s \cite{fchir},\cite{raff3} for the processes
$ \nu_e+\bar \nu_e\rightarrow \nu_x+\bar \nu_x$, in which the initial state is so simple that there is no need for the four beams indexing with ${j,k}$; there is already an equation for each beam in each of the sets (\ref{array1}) and (\ref{array2}).

\subsection{2. A problem with the literature}

It appears to be the accepted dictum that early universe $\nu$'s that begin exactly in the equilibrium un-mixed flavor configuration can suffer no effects from the fast $\nu-\nu$ interaction. This in conflict to the basic results of 
 ref. \cite{RFS1}.  It is possible that the cause of this disagreement is a widespread misunderstanding of the initial conditions that are required when we deal with thermal ensembles.
 
As explained above, a $\vec \sigma$ operator is built to be applied to composite states built of a very narrow-coned swarm of 
N $\nu$'s of flavor A, say, as,
 \begin{eqnarray}
| \Psi_A\rangle= e^{i \sum_j \phi_j} \prod_{j= 1}^{N}| {\rm flav}_{j}^A \rangle \,,
\label{inter1}
\end{eqnarray}
where we explicitly put in the unknowable phase factor that comes with every $\nu$. 

 We expect this phase factor, which does not change during a period of coherent evolution, not to affect final results.   
 When we deal with the collectivization of the wave-functions within our very narrow cone in the way that matches
the corresponding collectivization, (\ref{coll}), in $H_{\rm eff}$, we must implement the step,
 
\begin{eqnarray}
 \prod_{j= 1}^{N}| {\rm flav_j^A}\rangle \rightarrow  |{\rm flav^A}\rangle
 \label{inter2}
 \end{eqnarray}
and the phase factor in (\ref{inter1}) then tags along mutiplying the whole expression.
Now suppose, e.g. that the states A and B are respective flavor choices for a right-moving $\nu_e$ beam where each $\nu$  carries $\sigma_3^j=1$, and a right-moving
$\nu_x$ with $\sigma_3^j=-1$, as in the ideal early-universe case.
 \begin{eqnarray}
| \Psi_B\rangle= e^{i \sum_j \phi_j'} \prod_{j= 1}^{N}| {\rm flav}_{j}^B \rangle \,.
\label{inter3}
\end{eqnarray}

Someone comes along and says ``but now we don't need so many beams, we can just add the two flavor variables together and use that as the initial flavor configuration for that part of the wave function."  giving,
\begin{eqnarray}
 | \Psi_A\rangle+| \Psi_B\rangle=e^{i \sum_j \phi_j} |{\rm flav^A} \rangle+e^{i \sum_j \phi'_j} |{\rm flav^B} \rangle
 \label{inter5} \,,
  \end{eqnarray}
  where we should have done,
  \begin{eqnarray}
 | \Psi_A\rangle| \Psi_B\rangle=e^{i \sum_j \phi_j} |{\rm flav^A} \rangle e^{i \sum_j \phi'_j} |{\rm flav^B} \rangle
 \label{inter6} \,.
  \end{eqnarray}
 In view of its dependence on phase factors that are random, (\ref{inter5}) cannot play a role in a correct calculation. Our initial amplitude in the present problem must be in the form of a product over all 16 of the [flavor]$\times$[lepton number]  states, with no superpositions allowed.

We mention a recent paper \cite{bd1} devoted to looking at fast flavor effects in the early universe that could result from introducing {\it ad hoc} anisotropic density variations in the medium, and monitoring repeated changes in sign of the electron lepton number (ELN) distribution that result from these inhomogeneities. A prerequisite for doing this calculation is the belief that the standard perfectly homogenious early universe is stable, over many fast neutrino instability times. In our present work as well as in \cite{RFS1} the {ELN} starts out as zero everywhere. The authors of  \cite{bd1} might have erroneously taken that as a sign that nothing happens. But it also appears to us that their formalism leads straight to the (\ref{inter5}) issue.

A larger concern pertains to a genre of publications that are intended to make difficult systems like the supernova flow more computable by using some other basis than plane waves, for example angular moments. But it would appear that abandoning a basis of plane waves in favor of one of moments always means additive superpositions of plane wave states. In normal problems one can say ``If we use a complete set of angular functions that are best adapted to the geometry of the flow we can always get the plane-wave behavior back." In the present situation, though, we would have created something analogous to (\ref{inter5}) as an admissible state, and it would have the same fatal flaw as that state. 
Plane waves are the required basis for ``fast" $\nu$ work precisely because neutrinos move in exact straight lines during fast processes, no matter what the angular distribution might be.

\subsection{3. The outcome and interpretation of a 16-beam approach  } We take each initial particle to be in a definite flavor and lepton number state, defined as a simultaneous  eigenstate of the operators $[\sigma_3,\bar \sigma_3, \tau_3, \bar \tau_3]=[\pm 1, \pm 1,\pm 1, \pm 1]$, because of its previous production in some scattering event. Then each initial beam, as defined earlier, has the same set of possible values after the redefinitions that remove the factor of N. 
What is not allowed are any other values than $\pm 1$. These would always land us in the pit exemplified in (\ref{inter5}) in which the phase factor intervenes to destroy all value, rather than as an innocuous multiplicative phase factor multiplying the entire amplitude. 

In ref.\cite{RFS1} we reported on results of computational programs for solving the generalizations of  (\ref{array1}),  (\ref{array2}) for all angles ( where the $\vec \sigma$,$\vec \tau$ variables are no longer appropriate).
Specializing these results to the case where we can take $\cos \theta=-1$, we show in fig.1 an example of the results for the case in which we begin with what would have been an exact equilibrium configuration (in the absence of coherent effects): equal numbers of each of the four species in both the R and L beams. Indeed if we just asked for the expectation value of any of the operators $\sigma_3,\bar \sigma_3,  \tau_3,  \bar \tau_3$, in a sum over all beams, they each would still remain zero. But we wish to look more deeply, and examine instead the flavor density of each of the 16 beams individually. And the reader must keep in mind that a beam is a set of momenta; it has definite flavor only at the single initial time, and in some cases at periodic recurrences.

In fig.1 the solid curve shows behavior in a channel of the R species that is initially in the $\nu_e$ mode and 
the dashed curve behavior in an R channel that is initially in the $\nu_x$ mode. In a half-cycle of the periodic oscillation we come back to a pure flavor case, but with these assignments reversed. Each of the other 14 amplitudes 
duplicates one of these plots or the other. Indeed, there are only a small regions in which there is much mode-mixing: where the amplitudes break from the plateaus  at $\pm 1$. The dashed lines bracket a region that we shall call the ``instantons".

\begin{figure}[h] %  figure placement: here, top, bottom, or page
 \centering
\includegraphics[width=2.5 in]{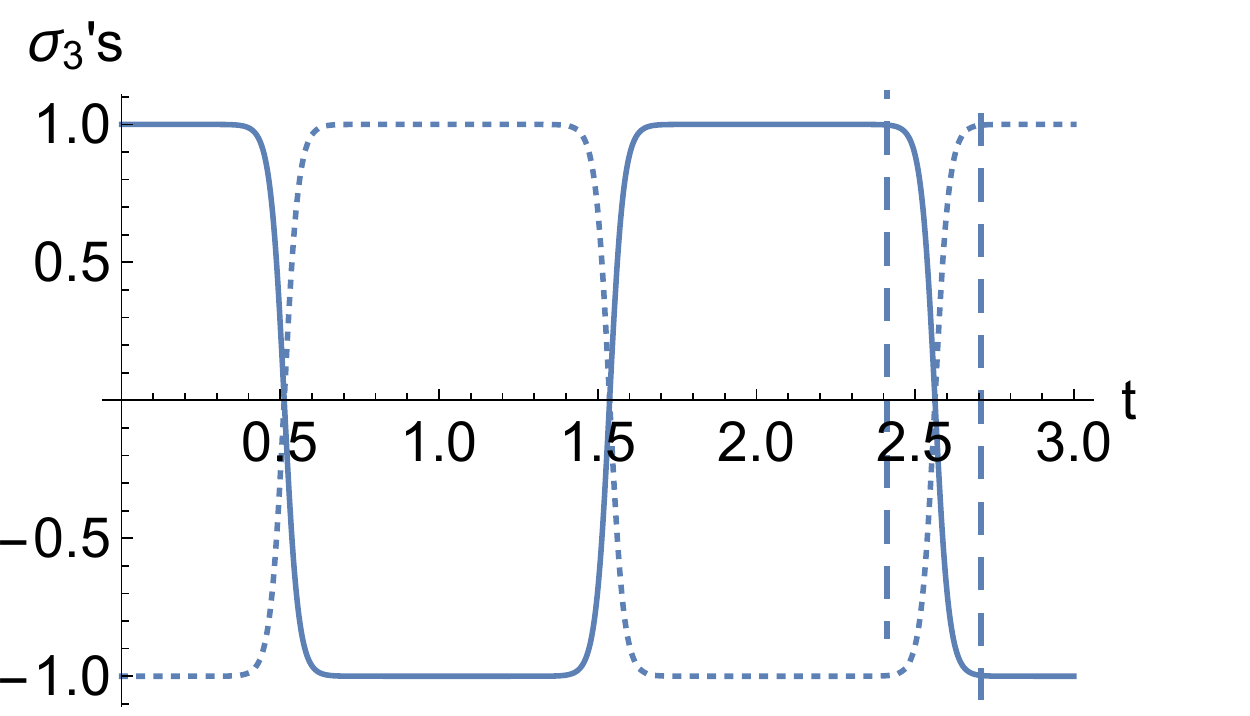}
 \caption{ \small } 
Solid curve: the variable $\zeta=\sigma_3=n_e-n_x$, where the $n$'s are occupation averages in the up-moving beam that began as pure $\nu_e$. Dotted curve: the same for the beam that began as pure $\nu_x$.  For a number density (of each species) of $(3~{\rm MeV})^3$ the time scale is in units of $ \hbar \, 10^{5} ({\rm eV} )^{-1}$,  approximately. \label{fig.1}
\end{figure}

 One feature of the instability mechanism calculation that we address here only in brief is that it does require seeds. None of the fast instabilities discussed in the present paper are seeded by the usual neutrino mass terms.  But they are seeded in a completely calculable way and are in no sense ``random quantum fluctuations". Their origin is in the fact that the Heisenberg equations 
 for $ \sigma_\pm$ etc. as given in (\ref{array1}) are still operator equations in principle.  Assuming that they hold for the expectations (``mean field") only picks up what in the end is the leading order in powers of $\hbar$. In fact, a better, but more complex, approach to deriving evolution equations from $H_{\rm eff}$ leaves behind seed terms with dependence $N^{-1}$. 
 A fuller treatment of the issue, but illustrated in a simpler model, is given in the supplementary material.
 The length of the plateau is then seen to be of order $G_F (n_\nu)^{-1} \log N$ while the length the dive remains of order $G_F (n_\nu)^{-1}$.  This general property is confirmed in complete solutions of a field theoretical model with 2000 states
  that corresponds in many qualitative ways with the theory here. See eq. 4  and fig. 1 of ref.\cite{rfsgrav}.
  
   \subsection{4. True instantons? And what about SU3?}
   
An early paper \cite{DS} looked at the simplest model of fast behavior in a toy model that left out ordinary flavor altogether. But it did contain both ``$\nu$" and ``$\bar\nu$" and found result's that might lead to a deeper understanding of the whole subject, namely,
 
 1)  Curves of time behavior as plotted for various values of N, all having the shapes of eigenfunctions in a symmetric quartic double-well potential.
  
  2).  These curves also being solutions of an equation, classical in appearance, with $d^2/(dt)^2$ on the LHS and the
 inverted potential, with twin peaks, on the RHS. 
 
 The latter behavior had also been noted in ref.\cite{RFS2}, eq.(23), in a multi-angle paper that included flavor.
  The curves shown in fig. 1 have precisely the same functional dependence, as closely as we can calculate it, to those found in the above. In the classical analogue they are ones
 that begin at rest a tiny bit to the right of the left-hand peak. This ``tiny bit" is related to the quantum terms that determine the ``plateau" length discussed above.
 
 This situation is exactly the same as described in, e.g., the introductory ``double-well" parts of Coleman's famous ``The Uses of Instantons" lectures \cite{ SC}. Is it too much to hope that some of the confusions complicating our lives will be avoided in a future formulation based on the gauge theory that gave birth to the Z's and their coupling to 
  $\nu$'s? One beautiful argument from  \cite{ SC} took the SU2 results from instanton considerations and extended them to SU3 simply through a topological argument. Is there such a way of avoiding the sweat and toil of the three-flavor problem in our case?

Coming back to earth, flavor SU3 for the neutrino triplet surely belongs in the picture when one extends to three flavors.  The effective Hamiltonian, an SU3 invariant, then is expressed in terms of the eight 3$\times$3 matrices $\lambda_j$  for the R direction, and a second set, $\hat\lambda_j$ commuting with the first, for the L direction. The internal commutation rules for both sets are as defined in the Wikipedia article, ``Clebsch-Gordan coefficients for SU3".  The $\vec \sigma \cdot \vec \tau$ form in our work gets replaced by $\sum_{i=1}^8 \lambda_i \hat \lambda_i$. Everything is parallel to the SU2 case except that now there will be four times as many equations in the analogues to (\ref{array1}), (\ref{array2}).
  
 \subsection{5. Calculation of the modified Dodelson-Widrow process.}
 A heavy (say mass=$m_s$=1 KeV) sterile $\nu_s$ that mixes with an active flavor through a small off-diagonal element in the 
 $\nu_s $ mass matrix is a long-standing contender as a dark matter candidate. We show here that the behavior of our individual beams, in our new construction of the amplitudes for the universe at $5 ~{\rm MeV}< T<10 ~{\rm MeV}$, drastically affects
 the previous results with respect to allowed parameter space for this process. We shall deal now with a single mode,
 $\vec q$, of the neutrino. For this mode we introduce annihilation operators $a$ and $b$ for the two flavors  
 (e, and x) of active $\nu$'s, and also introduce $s$ as the annihilation operator for the sterile,
 \begin{eqnarray}
 \nu_e\rightarrow (a^\dagger,a) ~,~\nu_x\rightarrow (b^\dagger,b) ~,~
 \nu_s\rightarrow (s^\dagger,s)  ~.
 \end{eqnarray}
The basic active-sterile coupling is,
  \begin{eqnarray}
 H_{A-S}=\lambda (a^\dagger s+s^\dagger a)+m_s s^\dagger s \,,
 \label{has}
\end{eqnarray}
and an ordinary active neutrino oscillation term is obtained by setting $g(t)=1$ in, 
  \begin{eqnarray}
 H_{\nu_{e}, \nu_{x}}=\lambda' g(t)(a^\dagger b+ a b^\dagger)\,.
  \label{bigmix}
 \end{eqnarray}
  To be consistent with data from other venues we are limited to active-sterile oscillations with an amplitude of about .01.
  And at a temperature of 10 MeV a $\nu$ has only a few collisions left in the time remaining before 
 $\nu$ decoupling. So collisional liberation cannot liberate many real steriles when the mixing is as small as .01. 

 Now we temporarily set $\lambda=0$ in (\ref{has}), and look for a function
 $g(t)$ which, when placed in (\ref{bigmix}), describing oscillation in the e-x sector,  produces a 
 $\sigma_3 (t) $ that is a fit to the one of the plots of fig.1. We find a function $g(t)$ that produces the plots of fig. 2, and judge that this function will suffice.
   \begin{figure}[h] %  figure placement: here, top, bottom, or page
 \centering
\includegraphics[width=2.5 in]{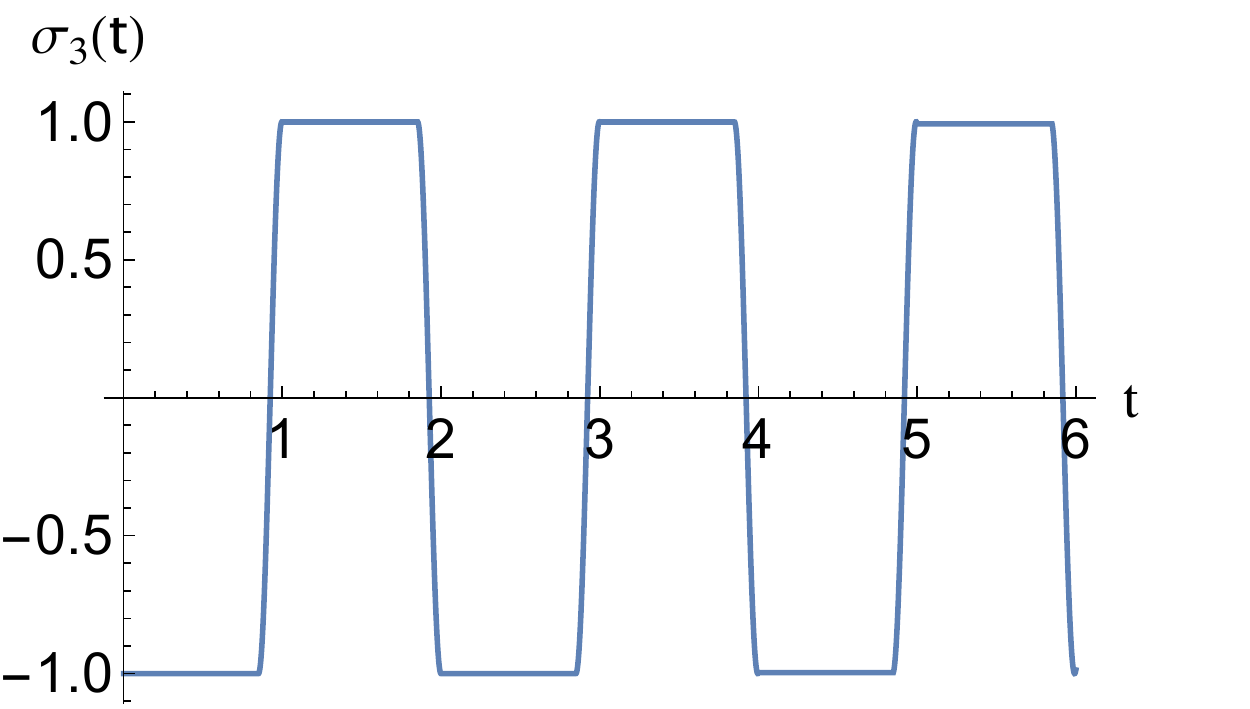}
 \caption{ \small } 
Plot of function ``$\sigma_3(t)$", that approximates that shown in fig.1.  But it is now produced by solving the equation for  $\sigma_3(t)$ generated from (\ref{bigmix}) in the case of no active-sterile coupling.  Taking a particular $g(t)$ it produces curves of the form shown in fig. 1, with their alternate plateaus and plunges, showing how the instanton plunges and ascents can be mocked-up in a one body simulation.
\label{fig.2}
\end{figure}

Using this function $g(t)$ we look afresh at the production of the sterile $\nu$ where the instanton interruption replaces scattering as the liberation mechanism. We restore the constant $\lambda=.1 ~{\rm eV}$ in (\ref{has}).  We calculate the flavor evolution for the single momentum, three flavor system fueled by
 (\ref{has}) and  (\ref{bigmix} )  and over a time range that corresponds to what would be six (tiny) active-sterile 
 oscillations in the absence of the surrounding medium. The results for the combined interactions are shown in fig. 3.,
 
 \begin{figure}[h] %  figure placement: here, top, bottom, or page
 \centering
\includegraphics[width=2.5 in]{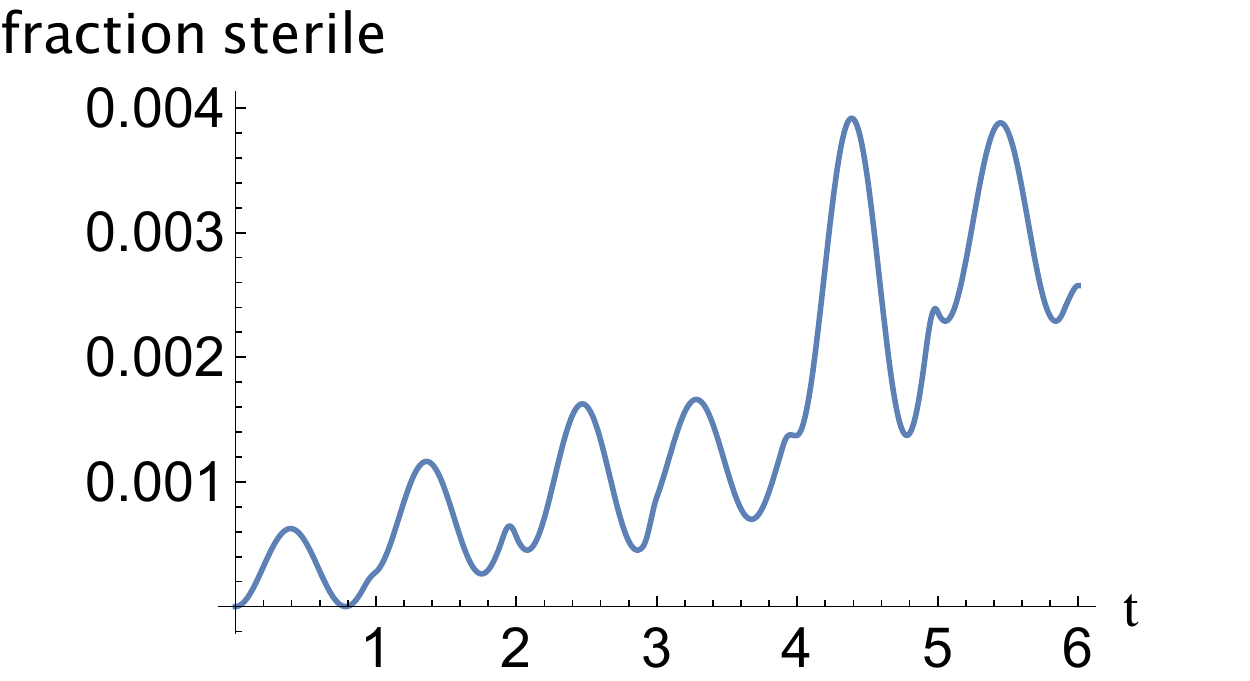}
 \caption{ \small } A time plot of the cumulative probability of the conversion of a single $\nu_e$ to a $\nu_s$ through the agencies discussed in text. The units of time are the same as in the plot of fig.1
 \label{fig.3}
\end{figure}
The successive narrow major peaks in the above plot more or less coincide in time with
the broad peaks in the instigating disturbance $g(t)$. Does the rise of the minima go on forever? Our numerical experiments tend to fall apart for times larger that those shown. But as to possible importance, we should bear in mind that at a temperature of 5 Mev the time scale for our entire plot  is of order 10$^9$ faster than
the collision time of these $\nu$'s with other $\nu$'s. Thus the way that we envision producing large numbers of steriles with small coupling $\lambda=10^{-3} m_s$ is through a number N$_{\rm I}$ of repeated short interactions. The probabilty of creation of a real sterile is now of order N$_{\rm I}\times$ 10$^{-6} $, rather than ${10}^{-6} $, Therefore copious sterile production appears to be predicted.

In apology for the crudity of the above model, we should say that a better approach would be letting our basic 
$H_{\rm eff}$ from (\ref{hammy}) contain a modest addition that has all modes of the $\nu_s$ canonically coupled
to the $\nu_e$ field (and $\bar \nu_s$ to  $\bar \nu_e$). We get just four more equations of motion, for each
direction, and everything appears contained, but our Mathematica program says ``too big" so the problem waits for bigger and stronger (and funded) groups to show up.
\subsection{6. Discussion}
The most important parts of this paper are not the details of the plotted results, but whatever leads to better understanding and application of the arguments discussed in section 2.  The actual 3D simulations mentioned in \cite{RFS1}, which were rough, have not been repeated. But the implications of section 2 of the present paper spell out the necessity of the 16-beam approach that was followed there. The theoretical supernova $\nu$ world,  which has generated a huge flow of papers recently, is certainly a domain in which the questions that we raise need to be considered. But we suggest that the early universe, a much simpler system, is perhaps the better for systematic investigation.
And it is more important to physics than supernova stuff, with or without the sterile $\nu$ ad-on.

On the positive side, there is the exciting new possibility of producing plentiful weakly coupled, KeV mass sterile $\nu$'s without invoking other hypothetical particles or couplings that are outside the standard model, as done in \cite{adg}.  
 In a universe without sterile $\nu$'s, there is another set of present understandings that need to be challenged.
These relate to the precision calculations of $\nu$ relic numbers and energy spectra, and of the often quoted ``effective number of $\nu$ flavors". The impact of the new collective $\nu$ results stems from the fact that even the smallest imbalance between $\nu_e+\bar \nu_e$ and $\nu_x+\bar \nu_x$ occupancy gets removed at the same fast rate of order $G_F n_{\nu}$. To know how important the changes will be has to await a complete repetition of the work summarized in the review article by  Dolgov,  Hansen, and Semikoz \cite{dol}, but with our effects included. The subject is discussed at greater length in ref. \cite{RFS1}. Some relevant work on methods, but applied to slightly unphysical models, is discussed in the supplementary materials of the present paper.

\section{Supplementary material}
\subsection{1. Illumination from a simpler theory.}
The simplest effective interaction that can produce fast flavor processes is just the first part of (\ref{hammy}),
\begin{eqnarray}
H_{\rm eff}={G_F\over {\rm Vol.} }\sum_{j,k}\Bigr [\sigma_+^j\tau_-^k +\sigma_-^j\tau_+^k 
+{\beta\over 2}\sigma_3^j \tau_3^k \Bigr ]\,,
\label{hammy2}
\end{eqnarray}
where the physical value of $\beta$ is unity. And for this case it is easy to see that for the case of $\beta=1$ 
and a single line in space, with beams in the two directions, there is no growing mode, or ``fast flavor"
action. However if we look in the same single line configuration at the cases in which $0\le\beta<1$, there is fast  exchange of two flavors for every value in the range. The case $\beta=0$ was  (mistakenly) presented as a physical result in ref.\cite{BRS}. The physical choice $\beta=1$ was then solved analytically in ref.\cite{FL}, which found no growing mode. ``Marginally stable" would have been the best description of that outcome. 

Choosing an instability-inducing value $\beta=1/2$ we have also carried out some 3D calculations in the model (\ref{hammy2}) with no $\bar \nu$'s, where a basic 1(+)D calculation demands only 4 rather than 16 beams. The results  are much more precise than the previous 3D results mentioned in \cite{RFS1} in their plots of  turnover in the 3D case and again show the same functional dependence on time, but with a time scale that is somewhat greater.

\subsection{2. Seeding in simpler examples.}

 ``Seeds" are a name for initial values of 
$\sigma_{\pm}$  $\tau_{\pm}$ for use with mean-field (MF) equations such as (\ref {array1}, (\ref {array2}). But, as long as we are careful about operator orders, we can instead use Heisenberg equations for more complex operators to go beyond the basic MF results. 
For the simplest case consider the $\beta=0$ case of the example in the last section,
\begin{eqnarray}
H_{\rm eff}=  {N G_F \over V}  \Bigr [\sigma_+ \tau_-+\,\sigma_- \tau _+
\Bigr ]\,.
\label{ham6}
\end{eqnarray}
where the $\sigma_\pm$, $\tau_\pm$ operators were defined just before eq.(2) of the main text. This interaction is not directly applicable to a neutrino problem, since it lacks both the term $\sigma_3\tau_3 /2$, which would have choked off the instability, and the anti-particle terms, which then restore the instability in the complete theory. Restoring both of those terms would lead to a much more complicated calculation below, but following the same lines.

We choose a new set of variables, 
\begin{eqnarray}
&X=\sigma_+\tau_- ~~, ~~Y=\sigma_- \sigma_+ + \tau_+ \tau_-   ~~,
\nonumber\\
&Z=(\sigma_3-\tau_3)/2
\end{eqnarray}

The quantity $\sigma_3+\tau_3$ does not change in time.  We shall always choose it as  $\sigma_3+\tau_3=0$. We obtain the Heisenberg equations based on the commutators with $H_{\rm eff}$ of (\ref{ham6}), 
\begin{eqnarray}
 {i V\over N G_F }\dot X=Z (Y +Z)\,,
 \label{extra}
\end{eqnarray}
\begin{eqnarray}
{i V\over N G_F } {\dot Y}=  2Z(X^\dagger- X)\,,
\end{eqnarray}
\begin{eqnarray}
 {i V\over N G_F } {\dot Z}=2(X-X^\dagger)\,.
\label{eom3}
\end{eqnarray}

%\begin{eqnarray}
%I {d\over dt}[\tau_- \tau_+ \,]=-\sigma_3 \sigma_+\tau_- -\sigma_ -\sigma_3 \tau_+= X \tau_3- \tau_3 %X^\dagger
%\end{eqnarray}

The $-Z^2$ term in (\ref{extra} )comes from a second commutation to get operators into a standard order; implicitly it carries an additional power of $\hbar$ and provides the seed for something to happen.
Next we do a rescaling in which each one of the operators $\vec \sigma, \vec \tau$ is redefined by extracting a factor of $N$, so that $x=X/N^2$, $y=Y/N^2$, $z=Z /N$ and at the same time defining $n=N/V$, the number density, and a scaled time variable, $s$, according to \newline $s=( G_F n)^{-1} t$, where $n$=the number density
$N/V$ of each beam.
 
The rescaled equations in the MMF approach are,
\begin{eqnarray}
i {d x \over ds} =zy-z^2/N \,,
\nonumber\\
i {d y \over ds} =2 z ( x^\dagger-x) \,,
\nonumber\\
i {d z \over ds} =2 (x-x^\dagger ) \,.
\label{mmf}
\end{eqnarray}
leading to solutions for z(t) that, throughout the first oscillation, fit ones in which we use the original forms for the equations for,
$\sigma_+,\tau_+,\sigma_3$, but with seeding $\sigma_+(0)=N^{-1}$.
The generalizations of the above discussion and calculation to the physical case with its 16 beams, is really arduous, but we believe straightforward.
\end{document}